# Generating a Performance Stochastic Model from UML Specifications

Ihab Sbeity[1], Leonardo Brenner[2] and Mohamed Dbouk[1]

[1] Lebanese University, Faculty of Sciences, Section I
Beirut - Lebanon

[2] LSIS Laboratory – Laboratoire des Sciences de l'information et des Systèmes
Marseille – France

**Abstract**
Since its initiation by Connie Smith, the process of Software Performance Engineering (SPE) is becoming a growing concern. The idea is to bring performance evaluation into the software design process. This suitable methodology allows software designers to determine the performance of software during design. Several approaches have been proposed to provide such techniques. Some of them propose to derive from a UML (Unified Modeling Language) model a performance model such as Stochastic Petri Net (SPN) or Stochastic process Algebra (SPA) models. Our work belongs to the same category. We propose to derive from a UML model a Stochastic Automata Network (SAN) in order to obtain performance predictions. Our approach is more flexible due to the SAN modularity and its high resemblance to UML' state-chart diagram.

## 1. Introduction

Quantitative analysis of software systems is being recognized as an important issue in the software development process. However, it is widely accepted that performance analysis techniques have suffered a lack of acceptance in the wider software design community. The most convincing reason is the reluctance of designers to learn the specialized formalism required by Markovian numerical solution techniques.

To encourage designers to incorporate performance analysis, a wave of explicit efforts got a serious attention in the last two decades [1, 2, 3, 4, 11, 12, 13, 16]. Since the initiation of the software performance engineering (SPE) methodology by Connie Smith [16], three large groups of researches in the area have been noticed. The first group aims at constructing performance based frameworks that are able to be used by designers. This is the case of formalisms such Hit [2]. However, these formalisms had not received a good attention from designers because of the wide evolution of other powerful and dominant specification and design techniques such as LOTOS, SDL, and more specifically UML. Thus, a second group of works [1, 9, 12] in SPE proposes to extend existing specification formalisms by introducing performance information such as time and probability distributions to these formalisms. In [1], a prototypic version of a program package is described, which takes a TSDL (Timed SDL) model as input and creates an internal representation of an equivalent Finite State Machine, so that validation and performance evaluation of TSDL models can be done automatically. In [12], Authors gave raise to stochastic LOTOS which modifies the semantics of LOTOS to allow performance information to be represented and performance results to be computed. Designers have to be careful with the new notations which mostly imply the formal power of the specification to be lost.

Extending the notation in order to support performance information in the Unified Modeling Language (UML), is a recent attempt to merge the most widely used object oriented design notations. It has been adopted by the industry body, the object Management Group (OMG), as a draft standard [9]. A significant effort is underway to complete and refine this draft. Again, an additional effort is required from designers to deal with the proposed framework.

Recent efforts such in [3, 13] also show the eligible need to incorporate performance analysis in the UML specification process. However, these works treat specific applications and they do not give a general demarche that can be useful with other applications.

A significant approach does appear in SPE which consists of generating from specification formalisms, more specifically UML, a performance model. Up to date works in this area have explored the possibilities for simulation of UML models [7], generation of queuing network models from UML deployment diagrams and collaboration diagrams. Moreover, mappings from UML to stochastic Petri net models (SPNs), more specifically Trivedi's SPNP [8] tool's variant of SPNs, and to stochastic process algebra models, particularly Hillston's PEPA [11], have been also developed. All of these show the potential of using UML's logical and behavioral notations to define the structure of performance models. Furthermore, the power of this approach results from the ability of the performance formalisms to represent models with large state space.

The purpose of this paper is to initiate a new methodology in the SPE area and it is similar to the approaches above in the sense that we propose to generate a performance model from UML specifications. Starting from a UML model, we suggest engendering a Stochastic Automata Network (SAN) model [10] which may be used to predicate performance for large systems. The SAN formalism is usually quite attractive when modeling a system with several parallel cooperative activities. An important advantage of the SAN formalism is that efficient numerical algorithms have been developed to compute stationary and transient measures [5, 15]. These algorithms take advantage of structured and modular definitions which allow the treatment of considerably large models. Another important advantage of the SAN formalism is the recent possibility of modeling and analyzing systems with (phase type) PH distributions [14] allowing to model deterministic activities. In addition, SAN permits to represent a system in modular way. A SAN model is a state-transition graph having a strong likeness with the UML state-chart diagram.
For these reasons, we believe that SAN is more than adequate to generate a performance model from a UML specification model.

This work opens the door to propose a more general demarche of generating a SAN model from a UML model. Here, we illustrate our approach in an informal way based on an example. The rest of the paper is structured as follows: Section 2 presents an informal definition of the SAN formalism. Section 3 considers how to exploit UML for performance analysis. This includes the case study of a chess game in order to show the direct use of UML. Section 4 explores how the UML model maps into SAN based on our case study. Some typical results obtained by solving the model are also presented. Section 5 concludes our paper and describes our ongoing works.

## 2. Stochastic Automata Network

Stochastic Automata Networks, SANs, were first proposed by Plateau in 1985 [10!]. The SAN formalism enables a complete system to be represented as a collection of interacting subsystems. Each subsystem is represented by an automaton which is simply a directed and labeled graph whose states are referred to as local states, being local to that subsystem, and whose edges, relating local states to one another, are labeled with probabilistic and event information. The different subsystems apply this label information to enable them to interact with each other and to coordinate their behavior.

The states of a SAN are defined by the Cartesian product of the local states of the automata and are called the global states of the SAN. Thus, a global state may be described by a vector whose ith component denotes the local state occupied by the ith automaton. The global state of a SAN is altered by the occurrence (referred to as the firing) of an event. Each event has a unique identifier and a firing rate. At any moment, multiple events may be enabled to fire (we shall also use the word fireable to describe events that are enabled): the one which actually fires is determined in a Markovian fashion, i.e., from the relative firing rates of those which are enabled. The firing of an event changes a given global source state into a global destination state. An event may be one of two different types. A local event causes a change in one automaton only, so that the global source and destination states differ in one component (local state) only. A synchronizing event, on the other hand, can cause more than one automaton to simultaneously change its state with the result that the global source and destination states may differ in multiple components. Indeed, each synchronizing event is associated with multiple automata and the occurrence of a synchronizing event forces all automata associated with the event to simultaneously change state in accordance with the dictates of this synchronizing event on each implicated automata. Naturally, a synchronizing event must be enabled in all of the automata on which it is defined before it can fire.

Transitions from one local state to another within a given automaton are not necessarily in one-to-one correspondence with events: several different events may occasion the same local transition. Furthermore, the firing of a single event may give rise to several possible destinations on leaving a local source state. In this case, routing probabilities must be associated with the different possible destinations. Routing probabilities may be omitted only if the firing of an event gives rise to a transition having a single destination. Also, automata may interact with one another by means of functional rates: the firing rate of any event may be expressed, not only as a constant value (a positive real number), but also as a function of the state of other automata. Functional rates are defined within a single automaton, even though their parameters involve the states of other automata.

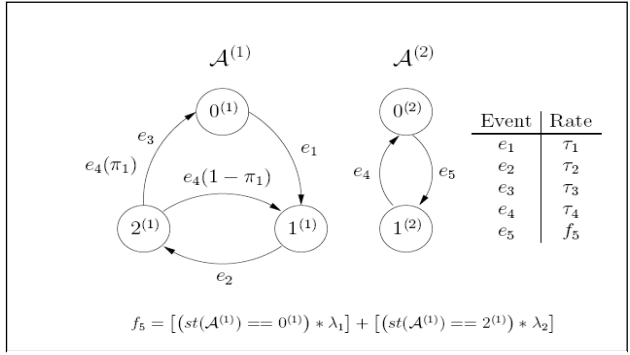

Fig 1: Example of a SAN model

As an example of the previous discussion, Figure 1 presents a SAN model with two automata, $A^{(1)}$ and $A^{(2)}$, the first with 3 local states, $0^{(1)}$, $1^{(1)}$ and $2^{(1)}$ and the second with two local states, $0^{(2)}$ and $1^{(2)}$. The model contains four local events, $e_1$; $e_2$; $e_3$ and $e_5$ and one synchronizing event, $e_4$. When automaton $A^{(1)}$ is in local state $0^{(1)}$, and $A^{(2)}$ is in local state $0^{(2)}$, (global state [0, 0]), two events are eligible to fire, namely $e_1$ and $e_5$. The event e1 fires at rate $\tau_1$. This is taken to mean that the random variable which describes the time $t$ from the moment that automaton $A^{(1)}$ moves into state $0^{(1)}$ until the event $e_1$ fires, taking it into state $1^{(1)}$, is exponentially distributed with a mean by $1/\tau_1$. Similar remarks hold for the firing rate of the other events. The firing of $e_1$ when the system is in global state [0, 0] moves it to global state [1, 0] in which $e_5$ is still eligible to fire, along now with event $e_2$. The event $e_1$ cannot fire from this new state. The synchronizing event $e_4$ is enabled in global state [2, 1] and when it fires it changes automaton $A^{(2)}$ from state $1^{(2)}$ to state $0^{(2)}$ while simultaneously changing automaton $A^{(1)}$ from state $2^{(1)}$ to either state $0^{(1)}$, with probability $\pi_1$, or to state 1(1) with probability 1-$\pi_1$. Observe that two events are associated with the same edge in automaton $A^{(1)}$, namely $e_3$ and $e_4$. If event $e_3$ fires, then the first automaton will change from state $2^{(1)}$ to $0^{(1)}$; if event $e_4$ fires the first automaton to change from state $2^{(1)}$ to either state $0^{(1)}$ or state $1^{(1)}$ as previously described. There is one functional rate, $f_5$, the rate at which event $e_5$ fires, defined as

$$f_5 = \begin{cases} \lambda_1 & \text{if } \mathcal{A}^{(1)} \text{ is in state } 0^{(1)} \\ 0 & \text{if } \mathcal{A}^{(1)} \text{ is in state } 1^{(1)} \\ \lambda_2 & \text{if } \mathcal{A}^{(1)} \text{ is in state } 2^{(1)} \end{cases}$$

Thus event e5, which changes the state of automaton $A^{(2)}$ from $0^{(2)}$ to $1^{(2)}$, fires at rate $\lambda_1$ if the first automaton is in state $0^{(1)}$ or at rate $\lambda_2$ if the first automaton is in state $2^{(1)}$.

The event $e_5$ is prohibited from firing if the first automaton is in state $1^{(1)}$.

$$f_5 = \left[(st(\mathcal{A}^{(1)} == 0^{(1)}) * \lambda_1\right] + \left[(st(\mathcal{A}^{(1)} == 2^{(1)}) * \lambda_2\right]$$

Functional transitions are written more compactly, e.g., in which conditions such as st($A^{(1)} == 2^{(1)}$) (which means \the state of $A^{(1)}$ is $2^{(1)}$) have the value 1 if the condition is true and are equal to 0 otherwise. This is the notation used to describe functions in the PEPS software tool [4]. In this setting, the interpretation of a function may be viewed as the evaluation of an expression in the C programming language. The use of functional expressions in SANs is not limited to the rates at which events occur; indeed, probabilities also may be expressed as functions. Figure 2 shows the equivalent Markov chain transition rate diagram for this example.

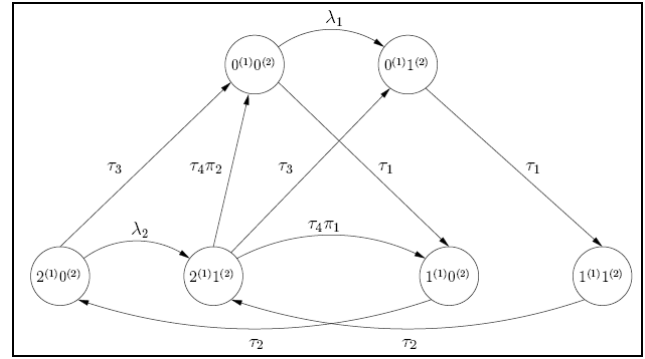

Fig 2: Transition rate diagram of corresponding Markov chain.

Furthermore, a new methodology has been recently incorporated into SANs: the use of phase-type distributions [14] The exponential distribution has been the only distribution used to model the passage of time in the evolution of the different San components. In [14], it is shown how phase-type distributions may be incorporated into SANs thereby providing the wherewithal by which arbitrary distributions can be used, which in turn leads to an improved ability for more accurately modeling numerous real phenomena.

The real interest in developing stochastic automata networks lies, in addition to their specification, to the fact that the *transition matrix* of the underlying Markov chain of a SAN can be represented compactly by storing the elementary matrices corresponding to subsystems and never the transition matrix itself. The numerical analysis of the system is then done by using the elementary matrices, which extremely decreases the analysis cost [5].

## 3. A UML model for the chess game

The Unified Modeling Language (UML) [9] is a graphically based notation, which is being developed by the Object Management Group as a standard means of describing software oriented designs. It contains several different types of diagram, which allow different aspects and properties of a system design to be expressed. Diagrams must be supplemented by textual and other descriptions to produce complete models. For example, a use case is really the description of what lies inside the ovals of a use case diagram, rather than just the diagram itself. For a full account of UML see [6].

Here, we develop UML model which shows how a chess game might look if based on object oriented design. We do not develop in detail all the possible UML diagrams, concentrating on those that are essential to the objective of describing the structure and behavior of the system. In particular, we only introduce the class diagram (section 3.1), collaboration diagram (section 3.2) and more specifically the state-chart diagram (section 3.3).

We do not spend time in presenting the use case model, which is important in real software design projects, since it captures the requirements for the system. We take those very much as given.

### 3.1 The class diagram

UML is an object oriented design formalism. Thus the core of the language is the *class diagram*. A class model defines the essential types of object available to build a system; each class is described by a rectangle with a name. This can be refined by adding compartments below the name which list the attributes and operations contained in each instance of (object derived from) this class [11].

Classes are linked by lines known as *associations* which indicate that one of the classes knows about the other. The direction of this knowledge is known as the *navigability* of the association. In an implementation an association typically corresponds to one class having a reference variable of the type of the other class. Sometimes navigability has to be two ways, but it more often one way. This can be shown by adding arrow head to the end(s) of the association.

For the purpose of this paper, we assume that classes and objects exist as fundamental units of description within a design. In particular, classes encapsulate behavior, which can be described as state machine description.

The class model of our chess game is reduced to its essentials. We assume there are two kind of players X and Y, the board and an umpire. A player spends time thinking before playing (achieving a movement on the board). Its movement may be valid or not. The umpire decides about the validity of the player movement.

Figure 3 shows classes for the chess game. We underline the existence of two classes (XPlayer and YPlayer) which inherit from the class Player. The inheritance relation is a form of generalization where one class is a super class; the other is a specification (sub class). The need to distinguish the two subclasses is related to the difference in behavior of the two players in term of state-transition. This is illustrated later in the state-chart diagram.

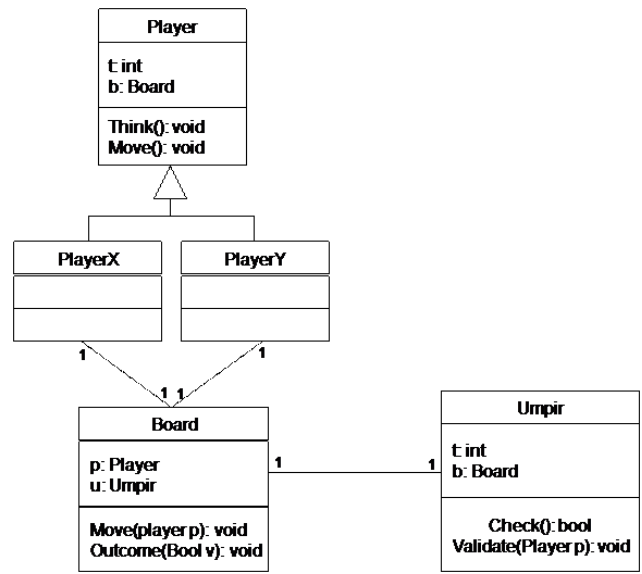

Fig 3: Chess game – The class diagram

### 3.2 The collaboration diagram

Collaborations are collections of objects, linked to show the relevant associations between their classes. Here "time" is not represented explicitly. Instead the emphasis is on showing which objects communicate with which others. Communications are represented by messages. Sometimes, these messages are numbered to show the order in which they should happen [8].

Figure 4 represents the collaboration diagram of our chess game. It acts of a communication between four objects: two players, *x* and *y*, aboard *b* and an umpire u. Also, we may for example imagine the existence of more than one player of type x and/or y that are playing on the same board. That implies the creation of new class instances which will be added to the collaboration diagram.

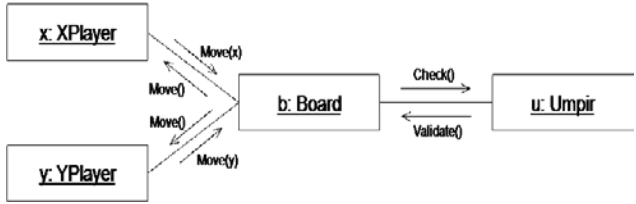

Fig 4: Chess game – the collaboration diagram

Collaboration (or communication) diagrams show a lot of the same information as sequence diagrams which are not represented here, but because of how the information is presented, some of it is easier to find in one diagram than the other. Communication diagrams show which elements each one interacts with better, but sequence diagrams show the order in which the interactions take place more clearly.

3.2 The state-chart diagram

UML defines state diagrams which allow a class to be defined in terms of states it can be in and the events (messages) which cause it to move between states.
Many software systems are event-driven, which means that they continuously wait for the occurrence of some external or internal event such as a mouse click, a button press, a time tick, or an arrival of a data packet. After recognizing the event, such systems react by performing the appropriate computation that may include manipulating the hardware or generating "soft" events that trigger other internal software components. (That's why event-driven systems are alternatively called reactive systems.) Once the event handling is complete, the system goes back to waiting for the next event.
State-charts describe how instances of classes behave internally. In a complete design they provide a full description of how the system works. We insert the state-chart for each object into its box in the collaboration. At any point in the lifetime of this system, each object must be in one, and only one, of its internal states. Each time a message (event) is passed, it may cause a state change in the receiving object and this may cause a further message to be passed between that object and another with which it has an association.
The overall state of a system will be the combination of all the current internal state of its objects, plus the current values of any relevant attributes. Intuitively, readers may sense a strong similarity to the stochastic automata networks behavior. An automaton is a set of states, transitions and events. The global state of a SAN is a combination of the local states of its automata. The powerful point of mapping the state-chart diagram into a SAN model is that the mapping process will not produce fundamental changes in the graph structure, only some information needed to represent relevant attributes and the time spent in a state is required. That may help designers to better understand the SAN performance model which is an emphasized advantage of our approach.

Figure 5 presents the state-chart diagram of our game model. For each object in the collaboration diagram, a chart is associated.

Each chart describes the internal behavior of the concerned object. Briefly, a chart is composed of states, transitions, triggers and actions. States are shown as lozenges; the initial state as a black filled circle. Transitions are the arrows between states, labeled with a trigger. Triggers represent the reason for an object to leave one state and follow the corresponding transition to another state; here, triggers are incoming messages or elapsing of time shown by the word *after* followed by the duration. Actions are resulting from a trigger carried out before entering the new state. They occurrence may follow a probability and/or they may involve sending messages to other objects, and these messages are prefixed by a caret. For example, in the automata corresponding to player X the label *after(t1)/^b.Move(x)* means that the action Move(x) of the Board b should occur when the trigger after(t1) is fired.

A player (X or Y) alternates between two states: the state where it is thinking about a move and the state where it is waiting its turn. Note that the initial state of X is thinking and that of Y is waiting (Player X starts the game). The two players cannot be simultaneously in the same state. When a player achieve a move, this move can be correct or not. The probability of a valid movement achieved by player X (respectively Y) is *p* (respectively *q*). The umpire alternates between two states: a state where it is checking a player's move and the state where it is idle. Parameters t1 and t2 are respectively the time needed by players X and Y to think about a move. The parameter T represents the time need by the umpire in order to validate a player's move.

## 4. Generating the SAN model

In this section, we present how a SAN model is directly generated for the UML chess game model presented in the previous section. The generation process is principally based on the state-chart diagram.

18

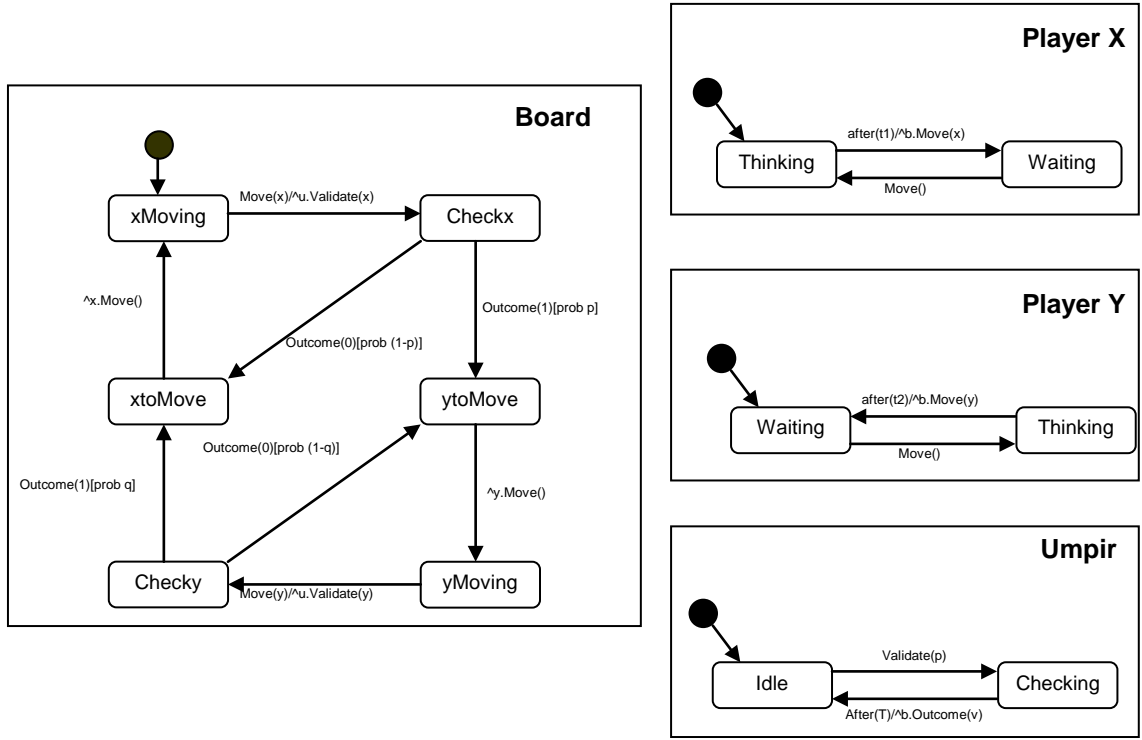

Fig 4: Chess game – the state-chart diagram

Even though the model of the chess game is not necessarily very significant from the performance point of view, but it is an excellent case study to show the simplicity of the generation process and consequently the significance of our methodology by using a SAN to generate a performance model from the UML specifications. Moreover, here, we focus only on the informal generation of the SAN model. The formal generation procedure is in the line of our future research activities.

The SAN model generated from the UML chess game model of section 3 is presented in figure 6.

As it can be noticed, the SAN model is robustly similar to the UML state-chart diagram of section 3.3. Each automaton of the state-chart diagram is mapped into a SAN automaton. But, in the general case, it is not necessarily that the number of the SAN automata is equal to that of the UML state-chart diagram. Sometimes, we need additional SAN automata in order to describe relevant attributes of the UML model. That is not the case of our chess game model. Moreover, each SAN automaton has a quasi-similar behavior to the state-chart, in term of states/transitions. SAN automaton X, Y, B and U correspond to UML's Player X, Player Y, Board and Umpire respectively (refer to figure 5). For all state-chart automata, the initial state (black filled circle) is omitted in the SAN automata, as there is no time spent in this state. The same time criterion is applied to states xtoMove and ytoMove of the Board state-chart automata. Again, there is no time spent in these states (we will call such states *discrete states*). Their use in the state-chart is needed only to model the consequence of a valid move achieved by a player that implies the second player to go to the thinking state.

States 0 and 1 of automata X and Y represent respectively *Thinking* and *Waiting* states of players X and Y. States 0 and 1 of automata B represent respectively states *Idle* and *Check*ing of the Board. States of automaton U (representing the Umpire) are obtained after the elimination of discrete states.

The generation of events is realized following a robust procedure: for each trigger corresponds an event. If the trigger does not fire an action in another automaton, the event is then a local event. Elsewhere, if the trigger fires an action in another automaton, the event is then a synchronizing event. If the fired action is a trigger for another action in another automaton, the same synchronizing event is used to synchronize all the affected automata. For example, looking to the state-chart of figure 5, the trigger *after(t1)* of Player X fires the action *Move(x)* of the Board which, in turn, fires the action *Validate(p)* of the Umpire. In the SAN model, this phenomena is represented by the synchronizing event "*a*" whose rate is equal to 1/t1. The SAN event "*b*" is generated in the same way as event "*a*".

On the other hand, the trigger *after(T)* of the state-chart Umpire fires the action outcome(v) of the Board. This action may have different parameters' value in the Board state-chart (v = 0 or 1). Thus, in principle, two events are needed to represent the firing of the trigger after(T). However, as the occurrence of outcome(0) and outcome (1) is related to probabilities, a new event decomposition is achieved in order to carry out the probabilities. Thus, the generated SAN events are now *u1q, u2q, u1p, u2p* that respectively correspond to *outcome(1)[prob q], outcome(0)[prob (1-q)], outcome(1)[prob p]* and *outcome(0)[prob (1-p)]*. The rate of the each SAN events is equal to 1/T times the corresponding probability. In addition, as the occurrence of the action *outcome* leads the state-chart Board to enter a discrete state, the action taken after leaving these discrete states (x.move() and y.move()) is taken into consideration in the SAN model as a consequence of the occurrence of *outcome*. This is due to the fact that the discrete states are not represented in the SAN model. That is why, for example, synchronizing event *u1q* appears also in the automata X (corresponding to the action x.move() fired in the state-chart Board).

Thus, the SAN model is generated following a procedure that can be easily implemented. On more detail remains to specify in the SAN model concerns values of time parameters, i.e. *t1*, *t2* and *T*. In addition, the type of the time distribution should also be indicated. SANs are basically Continuous Time Markovian (CTM) formalism. In CTM models, the exponential distribution is usually used taking advantage from its memory-less property which respects Markov theory. However, other time distributions may be also used in the SAN model [14] based on Phase-Type distributions that can approximate any time distribution even deterministic.

As a typical result, we present one performance prediction resulting from the SAN model with exponential distribution. The analyzed performance metric is the average playing time of each players (X and Y) when the parameters values are: p = q = 0.5; t1 = 2; t2 = 3 and T =1. The reflection time of player X is less than that of Y (t1<t2), the obtained result indicated that the total thinking time of player Y (over the whole time of the game) is 10% more than the thinking time of player X.

The goal from this simple result is just to show the adequateness of our methodology from mapping the UML model to a SAN model. Of course, more sophisticated performance analysis may be applied to more significant and large applications. In summary, our methodology performs extremely well.

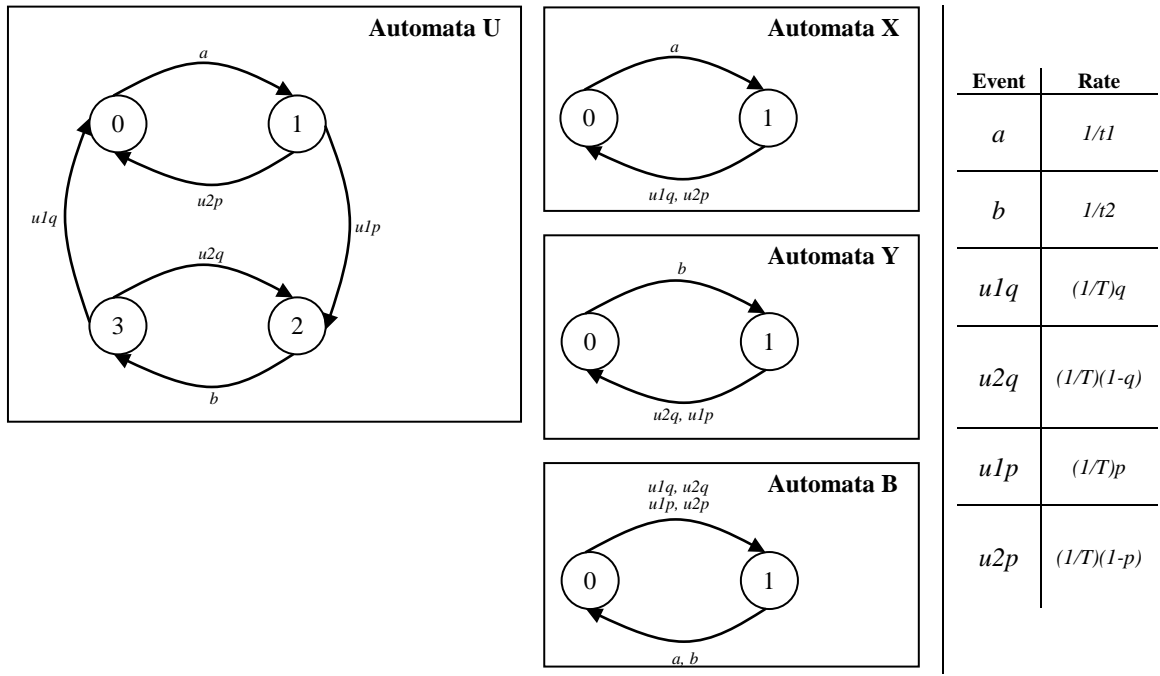

Fig 5: Generated SAN model

# 5. Conclusion

In this paper we have shown how a stochastic automata network may be directly generated from UML specifications. The key step is by using the state-chart diagram of UML which has a strong similarity to a SAN. Both, SAN and UML state-chart, are a set of automata. SAN's transitions are labeled by local events or/and by synchronizing events. UML's transitions are labeled by triggers that may fire actions in other automata. Mapping a state-chart automaton into SAN automaton is realized after eliminating UML discrete states. Mapping triggers and actions into events is achieved by following the path of actions carried out after the firing of a trigger. An informal description of the SAN generation procedure was presented based on a simple case study which goal is to show the simplicity and robustness of our demarche.

The methodology which we propose in this paper lays the groundwork for the development of a formal heuristic that permits to generate a SAN model from any UML state-chart model. In addition, the generation process is completely transparent to applications' designers. This opens the way for an efficient technique in the software performance engineering area by taking advantage of the power of SAN in analyzing large applications, SAN modularity, and its similarity to the UML state-chart diagram that may allow designers to easily interact with their performance model.

**Ihab Sbeity** – received a Maîtrise in applied mathematics from the Lebanese university in 2002, a Master in computer science - systems and communications from the "Université Joseph Fourier" – France in 2003, and a PhD from "Institut National Polytechnique de Grenoble", France in 2006. His PhD work is related to Performance Evaluation and System Design. Currently, Dr. Sbeity occupies a full time position at the Lebanese university – Faculty of Sciences I – computer sciences department. His research interests include modeling and performance evaluation of parallel and distributed computer systems, numerical solution and simulation of large Markov models, UML modeling and Software Performance Engineering (SPE).

**Leonardo Brenner** - Leonardo Brenner is assistant professor of Computer Science at Université de Provence, France, and do his research at Laboratoire des Sciences d'Information et des Systèmes. He got his Ph.D. (2009) in Computer Science degree from Institut Polytechnique de Grenoble (Grenoble INP), France. His research interests include performance evaluation of computer and communication systems, as well as theory and numerical solution of stochastic modeling formalisms for large Markovian models. His current research topics include structured formalisms, in particular stochastic automata networks and Petri nets, and also practical applications of performance and reliability modeling.

**Mohamed Dbouk** - Bachelor's Honor" in Applied Mathematics; Computer Science, Lebanese university, Faculty of Sciences (I)-

Beirut, PhD from Paris-Sud 11 University (Orsay-France), 1997. Specialty: Software Engineering and Information systems, Performance modeling and optimization, Geographic Information System & Hypermedia, Datawherehousing and Data mining.

He is a full time Associate Professor, Lebanese university - Faculty of Sciences (I) - Department of Computer Science. Academic Teaching: Database (including advanced topics), Object Orientation, Software engineering, Information System & Software Architectural Design, Distributed Architecture, Web Application Architectural Design. Research and Publications in: GIS, Cooperative and multi-agent systems, interoperability, 3D Modeling, Groupware, Graphical user interface.